\documentclass[10pt,sigconf,letterpaper,nonacm]{acmart}

\usepackage{graphicx}
\usepackage{subfigure}
\usepackage{amsbsy}
\usepackage{ulem}
\usepackage{upgreek}
\usepackage{algorithm}
\usepackage{algorithmicx}
\usepackage{algpseudocode}
\usepackage{threeparttable}
\usepackage{array}
\usepackage{caption}
\usepackage{url}
\usepackage{epstopdf}
\usepackage{verbatim}
\usepackage{color}
\usepackage{float}
\usepackage{wrapfig}
\usepackage{xspace}

\usepackage{enumerate}

\newcommand*{\etc}{
    \@ifnextchar{.}
        {etc}
        {etc.\@\xspace}
}
\graphicspath{{figures/}}

\begin{document}

\newtheorem{thm}{Theorem}
\newtheorem{lma}{Lemma}
\newtheorem{defi}{Definition}
\newtheorem{proper}{Property}

\title{\vspace{-1cm}Efficient Embedding VNFs in 5G Network Slicing:\\ A Deep Reinforcement Learning Approach}

\author{Linh Le,\hspace{0.2em}Tu N. Nguyen,\hspace{0.2em}Kun Suo,\hspace{0.2em}and Jing (Selena) He}
\affiliation{
\institution{Kennesaw State University, \hspace{0.3em}Marietta, GA 30060, USA}
\country{\{linh.le, tu.nguyen, ksuo, selena.he\}@kennesaw.edu}
}
\authornote{Corresponding author: Tu N. Nguyen}

\begin{abstract}
5G radio access network (RAN) slicing aims to logically split an infrastructure into a set of self-contained
programmable {\it RAN slices}, with each slice built on top of the underlying physical RAN (substrate) is a separate logical mobile network, which delivers a set of services with similar characteristics. Each RAN slice is constituted by various virtual network functions (VNFs) distributed geographically in numerous substrate nodes.
A key challenge in building a robust RAN slicing is, therefore, designing a RAN slicing (RS)-configuration scheme that can utilize information such as resource availability in substrate networks as well as the interdependent relationships among slices to map (embed) VNFs onto live substrate nodes.
With such motivation, we propose a machine-learning-powered RAN slicing scheme that aims to accommodate maximum numbers of slices (a set of connected Virtual Network Functions - VNFs) within a given request set. More specifically, we present a deep reinforcement scheme that is called Deep Allocation Agent (DAA). In short, DAA utilizes an empirically designed deep neural network that observes the current states of the substrate network and the requested slices to schedule the slices of which VNFs are then mapped to substrate nodes using an optimization algorithm. DAA is trained towards the goal of maximizing the number of accommodated slices in the given set by using an explicitly designed reward function. Our experiment study shows that, on average, DAA is able to maintain a rate of successfully routed slices above 80\% in a resource-limited substrate network, and about 60\% in extreme conditions, i.e., the available resources are much less than the demands.
\end{abstract}


\maketitle

\section{Introduction} \label{sec:intro}

The key aspect of RAN slicing is that the role of the {\it mobile network operator(s)} (MNOs) is to {\it coordinate} and allocate resources of the substrate network to ensure the harmonic coexistence of multiple RAN slices, while the role of the enterprise is to place slice requests and then manage the provided slices \cite{7926920,8004165,8057230,8004168}.
In particular, a RAN slice that is independent from others consists of a set of VNFs. Upon receiving requests from the enterprise, the MNO works with the RAN enforcement to allocate  the substrate network resources (e.g., resource blocks in long-term evolution (LTE)) to VNFs and virtual links between VNFs, and provide an appropriate mapping plan, according to the proffered demand, RAN policy, connectivity, and available spectrum resources of the substrate nodes, also called {\it resource blocks} (RBs).

There have been efforts to design algorithms for the RAN resource allocation problem that tend to consider only the available resources of the substrate network to design a mapping plan and disregard the VNF connectivity and bandwidth requirements \cite{8717789,7946928,6449268,doro2019slice,9071984,7529130,7417376,8377187}.
However, mapping VNF strategy will not only hinge on substrate nodes resource allocation but also rely on the substrate connection and the bandwidth requirement of the virtual links between VNFs. This is a key research challenge we will tackle in the proposed research.
In particular, for an RAN slicing (RS)-configuration in a cellular network, we would like to explicitly account for the {\it mapping plan} of VNFs, with the goal of providing {\it stable} performance for RAN slicing-based applications.

With such motivation, in this paper, we present a deep reinforcement scheme that is called \textit{Deep Allocation Agent (DAA)} that is able to embed VNFs of a slice while being aware of the interdependency among slices. We start with modeling the problem of maximizing the number of network slices to be fully accommodated as a reinforcement learning problem with the following settings: \textbf{1) Environment State} includes information on the current substrate network, i.e., available resources in all substrate nodes, and a set of pending  slices to be accommodated, i.e., VNFs in each slice and their demands. \textbf{2) Episodes} are the accommodation of all slices in a given set. An episode ends by allocating resources to all slices in the given set, or by entering an environment state in which no pending slices can be further resolved. Each step in an episode refers to the allocation of a slice. \textbf{3) Actions} that the agent makes at each step include selecting then allocating resources to VNFs in the slice. \textbf{4) Rewards} that the agent receives after an action are designed to guide the agent to generate schedule that maximize the number of allocated slices in a given set.

DAA then solves the problem of using a combination of a deep neural network and an resource allocation algorithm. The neural network first observes the current environment state to select a pending slice to accommodate. VNFs in then mapped to substrate nodes using an algorithm that maximizes the flexibility of the substrate network and the allocability of the remaining VNFs. We train the deep network of DAA as a deep reward network (DRN). In this case, the neural network predicts the true reward obtained if a slice is selected. The training of DRN is guided by the previously mentioned reward function. Our experiment study shows that, on average, DAA models are able to maintain a rate of successfully accommodated slices about 80\% in a resource-scarce case (i.e., the available resources are less than the demands from the slices), and approximately 60\% in extreme conditions of network (available resources are much less than demands). Specifically, the paper has the following \textbf{contributions} and \textbf{intellectual merits}:

\begin{enumerate}
    \item We tackle the problem of maximizing the number of slices in network slicing using deep reinforcement learning. Specifically, we propose a reinforcement learning model with specific designs of environments, actions, and rewards, that guide the agents towards a schedule that fulfills the most slices in a given set.
    \item We present DAA, a deep reinforcement  scheme that consists of an empirically designed deep neural network that schedules slices and an allocation algorithm that maps VNFs to substrate nodes while being aware of the network's flexibility and the VNFs' allocability. The deep network of DAA is trained either as a deep reward network to maximize the number of accommodated slices. DAA is shown to obtain good slice-resolving rates even in resource-scarce substrate networks, and is quickly to converge during training.
\end{enumerate}

\noindent \textbf{Organization:} The rest of the paper is organized as follows. Section $\S$\ref{sec:problem} discusses the concepts related to our work, and mathematically formalizes our research problem. We discuss DAA in Section $\S$\ref{sec:method} then present the experiment study in Section $\S$\ref{sec:exp}. Finally, we conclude our paper in Section $\S$\ref{sec:conc} and discuss future application in $\S$\ref{sec:dis}.

\section{Network Model and Research Problem} \label{sec:problem}
In the following sections, an overview of the quantum network and basic mathematical notations are introduced. Formal research problem is also defined to demonstrate the key ideas behind the proposed model and methodology.

\subsection{Network Model} \label{sec:def}

Given a substrate network $\mathcal{G}=\{\mathcal{N}, \mathcal{R}\}$ in which we have a set of nodes $\mathcal{N} = \{\mathcal{N}_1, \mathcal{N}_2, \dots, \mathcal{N}_n\}$ and a set of resources $\mathcal{R} = \{\mathcal{R}_1, \mathcal{R}_2, \dots, \mathcal{R}_n\}$. A node $\mathcal{N}_i$ has maximum resources available of $\mathcal{R}_i$. In other words, any node $\mathcal{N}_i$ can allocate a maximum amount $\mathcal{R}_i$ of resources to VNFs. A VNF in a specific slice is not available and accessible by other network slices for the isolation purpose.

Let $G = \{G^1, G^2, \dots, G^l\}$ be the set of $l$ RAN slices of which resources need to be allocated in $\mathcal{G}$. Each slice $G^i = \{V^i, R^i\}$ consists of a VNF set $V^i=\{v^i_1, v^i_2, \dots, v^i_{l_i}\}$ and a resource set $R^i=\{r^i_1, r^i_2, \dots, r^i_{l_i}\}$ in which the demanded resources for VNF $v^i_j$ is $r^i_j$. In other words, $v^i_j$ needs to be assigned to a substrate node $\mathcal{N}_k$ with $\mathcal{R}_k \geq r^i_j$ to be considered successfully mapped. To mathematicize the allocation of a VNF $v^i_j$ to a substrate node $\mathcal{N}_k$, we introduce a binary variable $m^{i,j}_k$. $m^{i,j}_k = 1$ when $v^i_j$ is mapped to $\mathcal{N}_k$, and $m^{i,j}_k = 0$ otherwise. A VNF can only be allocated to one substrate node, therefore we have the constraint $\sum_{k=1}^n m^{i,j}_k = 1$. The maximum resources available in each substrate node cannot be exceeded when allocating VNFs, which leads to another constraint $\sum_{i=1,j=1}^{l,l_i} m^{i,j}_k*r^i_j \leq \mathcal{R}_k$.
We further assume the slice $G^i$ to be successfully allocated only when all $v^i_j \in V^i$ are successfully allocated. Consequently, we introduce a binary variable $A^i$. $A^i=1$ when $\sum_{j=1,k=1}^{l_i,n} m^{i,j}_k = l_i$, and $A^i=0$ otherwise. In other words, $A^i=1$ when slice $G^i$ has all its VNFs successfully mapped to the substrate network.

\subsection{Problem Formulation}
\label{sec:prob}

In this work, we advocate a novel allocation scheme $-$ dubbed \textit{Deep Reinforce Allocation Scheme ($\mathcal{Q_{AS}}$)} $-$ that implements a deep reinforcement learning model to circumvent the computability limitations of \textit{heuristic} conventional resource allocation schemes, while concurrently maximizing serviceability of the network. In other words, our allocation scheme aims to maximize the number of network slices that are successfully allocated with resources to operate properly.

Given $G = \{G^1, G^2, \dots, G^l\}$ be the set of $l$ RAN slices of which resources need to be allocated in $\mathcal{G}$. Our goal is to maximize $\mathcal{Q_{RS}}$ as follows:
\begin{equation}\label{eq:obj}
\mathcal{Q_{AS}}=\underset{i \in \{1, 2, \dots, l\}} {\mathop{\max }}\,\sum{A^i}
\end{equation}
We further constraint the number of VNFs in all slices in $G$ to be equal $l_i = s \; \forall \; i$. It should be noted that this constraint does not reduce the generality of the research problem, since a slice $j$ VNFs with $j < s$ can be padded with VNFs of $0$ demands to meet the size requirement. In this problem, we also do not constraint that all slices in $G$ can be allocated successfully in $\mathcal{G}$. This means that a complete allocation solution may not exist for a mapping from $G$ to $\mathcal{G}$.

\begin{figure}[t]
\center
    \subfigure[Annotation]{\includegraphics[width=0.49\columnwidth, clip, trim=0cm 6cm 10cm 0cm]{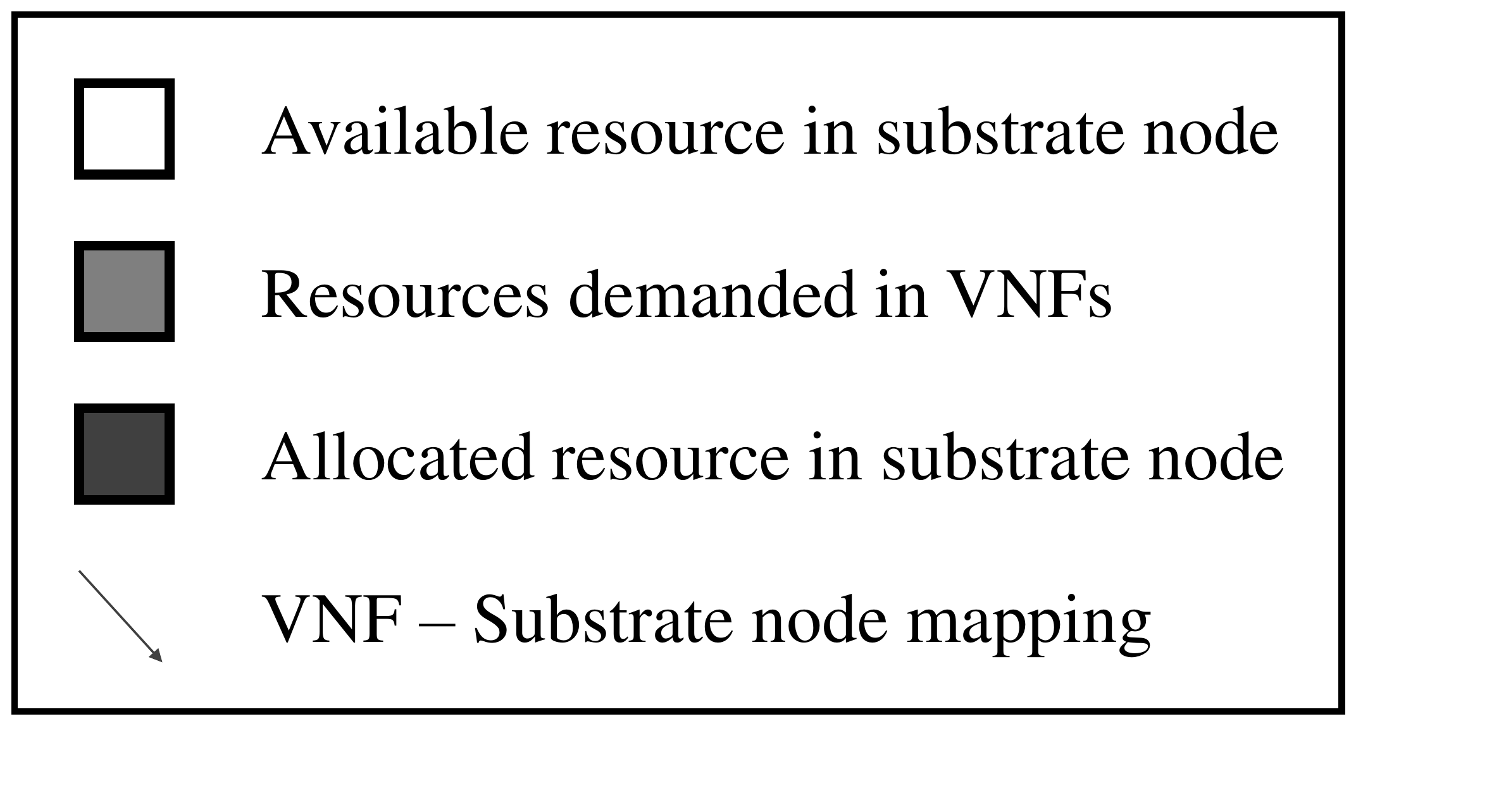}}
    \subfigure[Original Network]{\includegraphics[width=0.49\columnwidth, clip, trim=0cm 1cm 1cm 0cm]{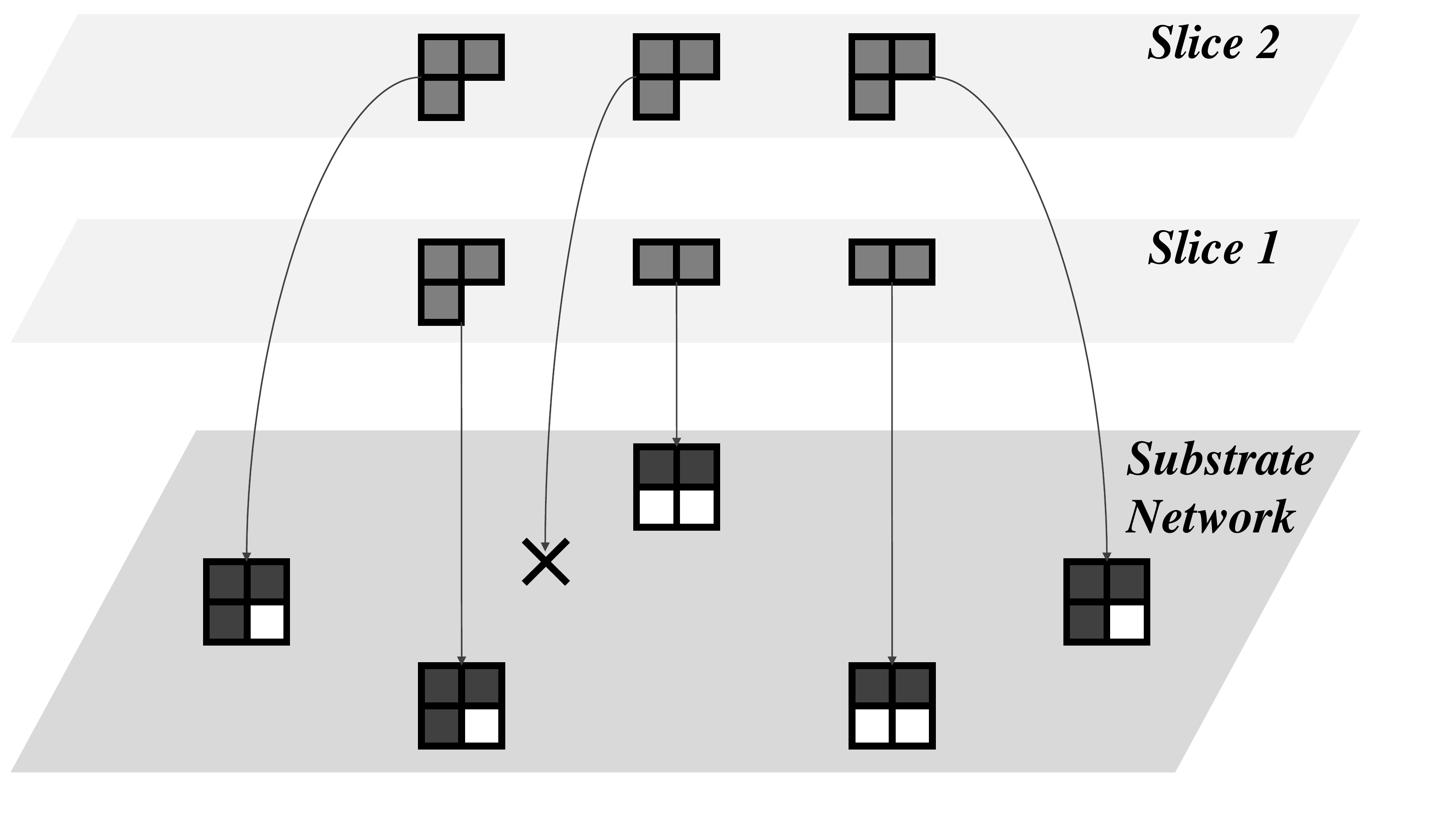}} \\
    \subfigure[Non-Optimal]{\includegraphics[width=0.49\columnwidth, clip, trim=0cm 1cm 1cm 0cm]{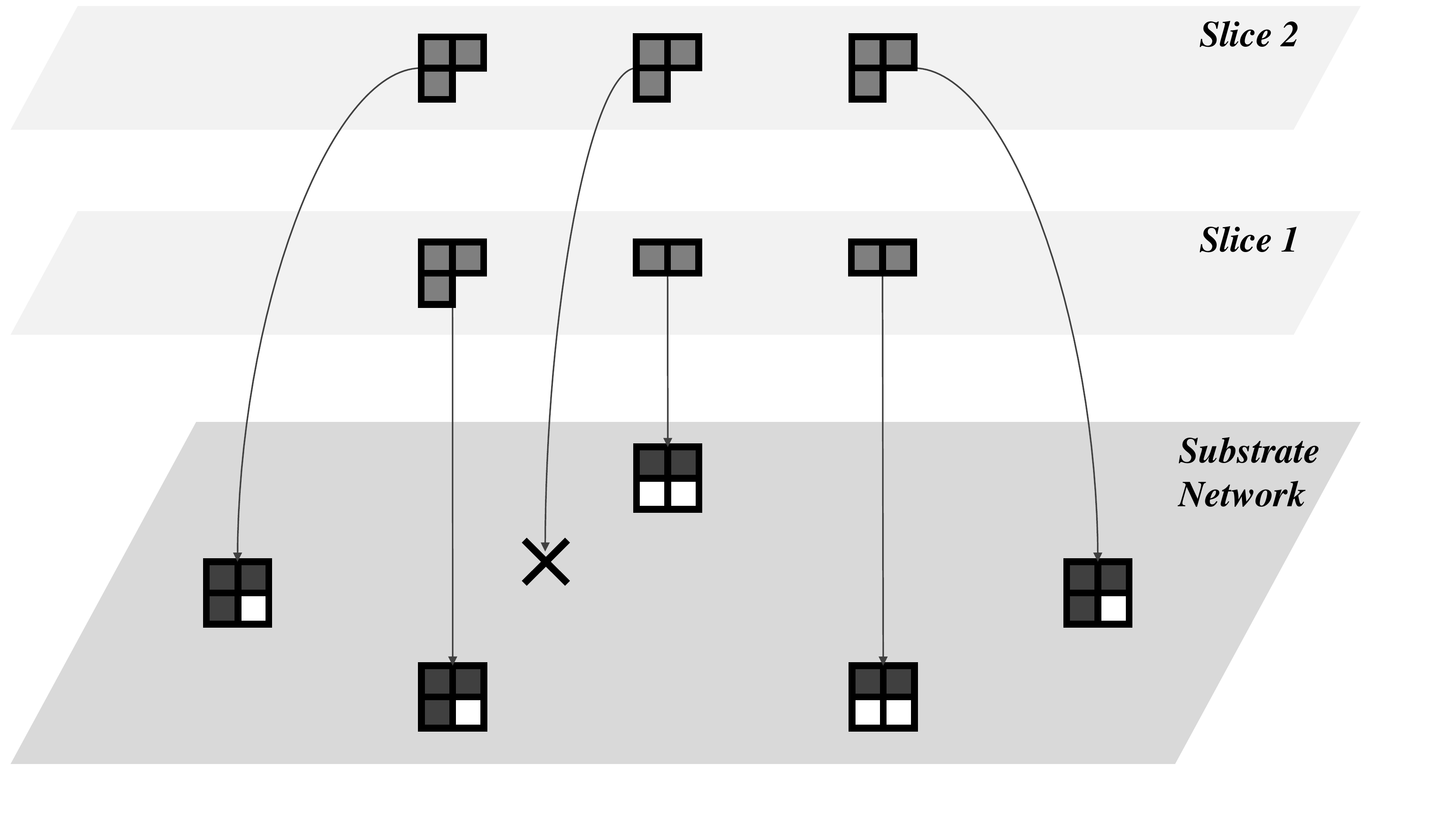}}
    \subfigure[Optimal]{\includegraphics[width=0.49\columnwidth, clip, trim=0cm 1cm 1cm 0cm]{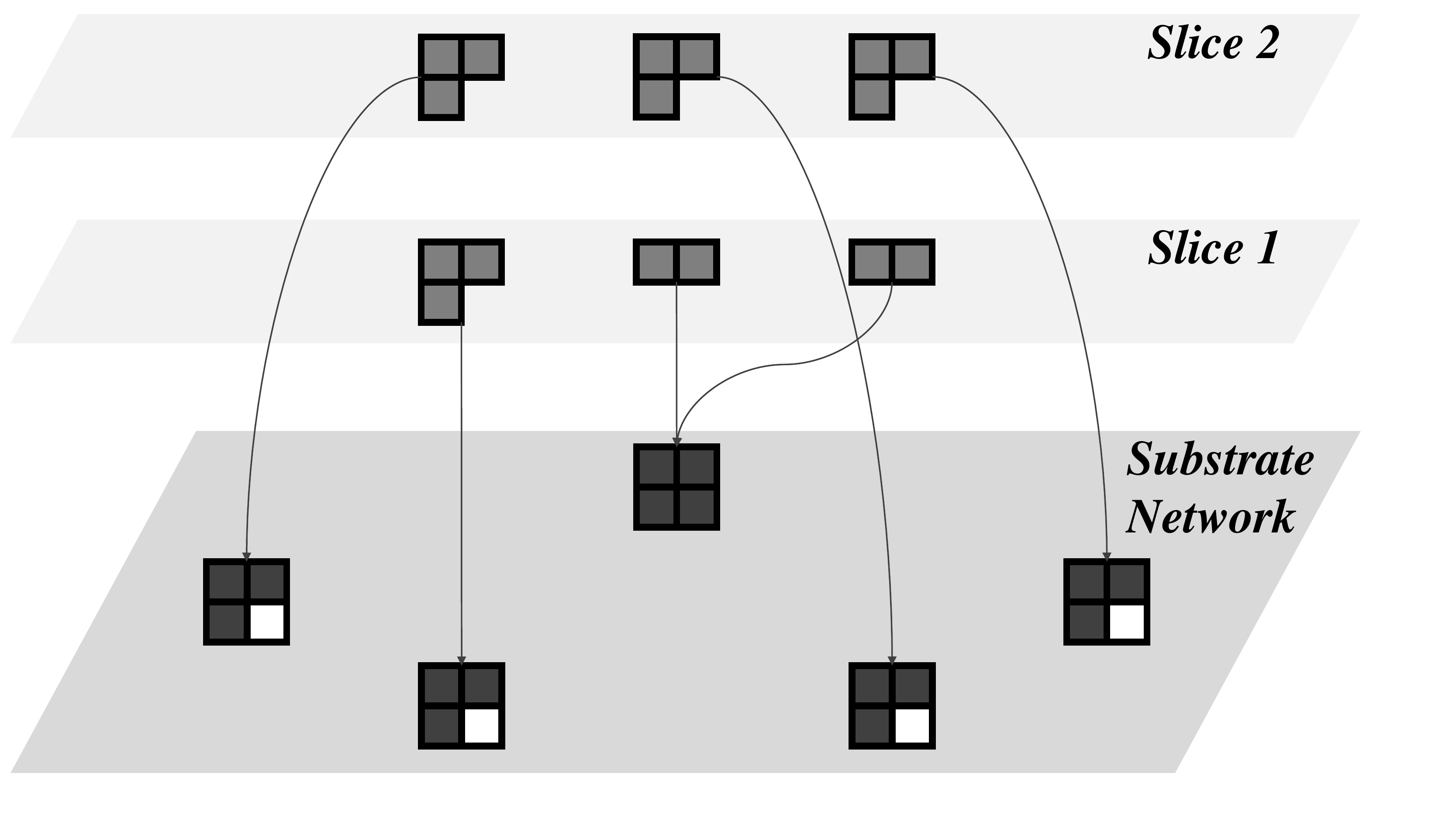}}
\caption{An example on a VNFs embedding problem and its non-optimal and optimal solutions}
\vspace{-15pt}
\label{Fig:example}
\end{figure}

We break the problem into two tasks which are 1) to schedule slices to allocate and 2) to map VNFs in the selected slice to substrate nodes. These two components make up the output of our $\mathcal{Q_{AS}}$ scheme. We shows an illustrative example in Fig. \ref{Fig:example}. Fig. \ref{Fig:example}(b) shows the substrate network $\mathcal{G}$ of five nodes, each with 4 resource blocks. There are two slices that need accommodations. Fig. \ref{Fig:example}(c) demonstrates the case in which slice 1 is accommodated first with the VNFs are randomly mapped to substrate nodes that satisfy the resource constraint. This solution is not optimal as slice 2 cannot be accommodated anymore. Fig. \ref{Fig:example}(d) shows the optimal solution in which slice 2 is accommodated first. Slice 1 can then be accommodated successfully as well. This example shows the importance of scheduling the RAN slices before allocating resources to them, as well as a mapping strategy that conserves resources based on future unallocated slices and VNFs.

\section{Deep Allocation Agent}
\label{sec:method}
In this section, we describe our deep reinforcement allocation scheme, henceforth referred to as Deep Allocation Agent (DAA). DAA utilizes a novel deep neural network to schedule slices and a VNF mapping algorithm to allocate resources to the selected RAN slice. The deep neural networks is referred to as a deep reward network. Given a substrate network $\mathcal{G}$ with the node set $\mathcal{N}$ and the resource set $\mathcal{R}$, and a set of RAN slices $G$, we refer to the accommodation of all slices $G^{i}$ in $G$ as an \textit{episode}. At each step $t$, DAA makes an action $a^{(t)}$ to allocate resources to \textit{one} slice $G^*$ based on the current environment state including the current available resources $\mathcal{R}^{(t)}$ and the demands of pending slices $R^{(t)}$ . An episode ends when all slices are resolved, or when no pending slices can be accommodated.
%
In the following sections, we first discuss our VNF mapping algorithm. Then, we describe the DNN architecture of DAA in terms of input states, output actions, reward function, architecture, and training.

\subsection{Allocating Individual VNFs' Resources}

The allocation of resources to a VNF, i.e., mapping a VNF to a substrate node, may significantly alter the performance of network slicing. Consequently, we derive mapping algorithm that select a substrate node for a given VNF while considering the \textbf{flexibility} of the substrate network and the \textbf{allocability} of the remaining VNFs. We first describe the concept of flexibility and allocability in the context of our problem.

\vspace{2pt}
\noindent \textbf{Substrate Network Flexibility.} We define the flexibility of a substrate network as the capability to accommodate pending VNFs. Mathematically,  given a substrate node $\mathcal{N}_k$ with available resource $\mathcal{R}^{(t)}_k$ at step $t$, and a RAN slice set $G$ with VNFs $\{v^i_j | i\in\{1,l\};j\in\{1,s\}\}$, the flexibility $\mathcal{F}(\mathcal{N}_k)$ of $\mathcal{N}_k$ is defined as\vspace{-10pt}
\begin{equation}
    \mathcal{F}(\mathcal{N}_k) = \sum_{i=1,j=1}^{l,s}\textbf{1}_{i,j}^k
\end{equation}
where $\textbf{1}_{i,j}^k$ is a binary indicator as follows
\begin{equation}
    \textbf{1}_{i,j}^k = \begin{cases}
        1 & r^i_j \leq \mathcal{R}_k \\
        0 & \text{otherwise}
        \end{cases}
\end{equation}

The flexibility $\mathcal{F^S}$ of the substrate network is then calculated as the sum of all substrate nodes' flexibility:
\begin{equation}
    \mathcal{F^S} = \sum_{k=1}^n \mathcal{F}(\mathcal{N}_k)
\end{equation}

\noindent \textbf{VNF Allocability.} We define the allocability of a VNF as the number of substrate nodes that can accommodate the VNF. Mathematically, the allocability $\mathcal{A}(v^i_j)$ of a VNF $v^i_j$ is defined as\vspace{-10pt}
\begin{equation}
    \mathcal{A}(v^i_j) = \sum_{k=1}^{n}\textbf{1}_{i,j}^k
\end{equation}
\vspace{2pt}
with $\textbf{1}_{i,j}^k$ defined previously. It can be seen that the lower $\mathcal{A}(v^i_j)$, the less substrate nodes can accommodate $v^i_j$, and when $\mathcal{A}(v^i_j)=0$, $v^i_j$, and subsequently $G^i$, are completely blocked from being successfully allocated with resources. Therefore, to minimize the possibility of a VNF and its slice being blocked, as well as further increase the possible solutions for future mappings of remaining VNFs, we will derive mapping solution based on the minimum allocability across all VNFs at each step $t$
\begin{equation}
    \mathcal{A^S}(v^i_j) = \min_{i,j}\mathcal{A}(v^i_j)
\end{equation}

\noindent \textbf{VNF Allocation Algorithm.} Given a VNF that needs resources, a substrate node is selected so that, \textbf{after} the allocation, both the substrate network flexibility $\mathcal{F^S}$ and the $\mathcal{A^S}(v^i_j)$ of the current network are maximized. More specifically, at step $t$, given a VNF $v^i_j$ of demanding resource $r^i_j$ and the substrate nodes $\mathcal{N} = \{\mathcal{N}_1, \mathcal{N}_2, \dots, \mathcal{N}_n\}$ with resources $\mathcal{R}^{(t)} = \{\mathcal{R}^{(t)}_1, \mathcal{R}^{(t)}_2, \dots, \mathcal{R}^{(t)}_n\}$, the process is as follows
\begin{enumerate}
    \item Temporarily allocate $v^i_j$ to each $\mathcal{N}_k$ that satisfies $\mathcal{R}_k \geq v^i_j$ and update $\mathcal{R}^{(t)}$ to obtain $\mathcal{R}^{(t+1)}_{(k)}$ with $\mathcal{R}^{(t+1)}_{k(k)} \leftarrow \mathcal{R}^{(t)}_k - r^i_j$. Calculate the flexibility $\mathcal{F^S}_{(k)}$ and allocability  $\mathcal{A^S}_{(k)}$
    \item Sort the substrate nodes $\mathcal{N}$ in descending order of $\mathcal{A^S}_{(k)}$ then $\mathcal{F^S}_{(k)}$ ($\mathcal{A^S}_{(k)}$ is prioritized)
    \item Select the first substrate node $\mathcal{N}_*$ in the sorted $\mathcal{N}$ to be the mapping solution for $v^i_j$
\end{enumerate}
This process ensure that $\mathcal{N}_*$ is selected so that the substrate network at $t+1$ has the maximum capability to accommodate the remaining VNFs, as well as has the lowest chance to block any VNFs and their slices completely.

\subsection{Deep Reward Network Architecture}

\vspace{3pt}
\noindent \textbf{Input State.}
At step $t$ in an episode $\mathcal{E}$, an input state to DAA consists of 1) the current available resource in each substrate node $\mathcal{R}^{(t)}$ and 2) the current demanding resources from all slices $R^{(t)}$. The data representations for $\mathcal{R}^{(t)}$ and $R^{(t)}$ are straightforward. $\mathcal{R}^{(t)}$ is modeled as a vector
\begin{equation}
    \mathcal{R}^{(t)} = \{\mathcal{R}^{(t)}_1, \mathcal{R}^{(t)}_2, \dots, \mathcal{R}^{(t)}_n\}
\end{equation}
whereas $R^{(t)}$ is modeled as a $l \times s$ matrix
\begin{equation}
    R^{(t)} = \begin{bmatrix}
        r^{1_{(t)}}_1 & r^{1_{(t)}}_2 & \dots & r^{1_{(t)}}_s \\
        \dots & \dots & \dots & \dots \\
        r^{l_{(t)}}_1 & r^{l_{(t)}}_2 & \dots & r^{l_{(t)}}_s \\
    \end{bmatrix}
\end{equation}
If a slice $G^i$ is already accommodated, its corresponding row $R^{(t)}_i$ in $R^{(t)}$ is replaced with a $\textbf{0}$ vector. Slices of less than $s$ VNFs can be padded with $0$'s to ensure the input matrix size.

\vspace{3pt}
\noindent \textbf{Output Action.}
At each step $t$ in an episode, DAA selects one among the pending slices $G^{(t)}$ then mapping its VNFs to the substrate network. We utilize a hybrid approach, that is to use the deep network SN for scheduling the slices, and the algorithm presented previously to map the VNFs.

In terms of scheduling, SN outputs a reward vector  $\rho^{(t)}$ of size $l$: $\rho^{(t)} = \{\rho^{(t)}_0, \rho^{(t)}_1, \dots, \rho^{(t)}_l\}$ in which $\rho^{(t)}_i$ represents the reward at the end of step $t$ if DAA selects slice $i$ to accommodate. Then, the \textit{pending} slice with the highest reward value is selected in each step, $a^{(t)} = \text{argmax}_{i\in G^{(t)}_P}(\rho^{(t)}_i)$, where $G^{(t)}_P$ is the set of pending slices at $t$. With $a^{(t)}$ selected, the VNFs in the chosen slice are sorted in descending order of demanding resources. Then, each VNF is mapped to the substrate network using algorithm that is described in Section $\S$3.A. The reason we first sort the VNFs in descending order of demanding resources is that, VNFs of higher demands will have less allocability and thus less mapping options. Therefore, they should be accommodated first to avoid being blocked.

At the end of each action, the input states are updated for the next step $t+1$. First, the row that associates to the slice resolved by $a^{(t)}$ in $R^{(t)}$ is replaced with $0$'s to obtain $R^{(t+1)}$. Then, $\mathcal{R}^{(t)}$ is updated with the new available resource values after accommodating all VNFs in the selected slice.

\vspace{3pt}
\noindent \textbf{Reward Function.}
The reward function is specifically designed to increase the number of accommodated slices by the substrate network. Mathematically, let the reward at step $t$ be $P^{(t)}$, then
\begin{equation}
    P^{(t)} = n^{(t)}_r\alpha + (l-n^{(t)}_r)\beta f + \lambda P^{(t+1)}(1-f)
\label{eq:reward}
\end{equation}
where $n^{(t)}_r$ is the number of slices that are resolved from the beginning of the episode through step $t$, $f$ is a binary indicator that is 1 when $t$ is the ending step of and episode and 0 otherwise, $\alpha > 0$ is the reward term, $\beta < 0$ is the penalty term, and $\lambda \in [0,1]$ is a discount factor. Both $\alpha$, $\beta$, and $\lambda$ are hyper-parameters to be selected during the training phase. Overall, the reward function design encourages the agent to find schedules that accommodate more slices, and penalizes schedules that end too early - the less slices solved, the heavier the penalties.

\begin{figure}[t]
\center
\includegraphics[width=0.7\columnwidth, clip, trim=0cm 5.5cm 16.5cm 0cm]{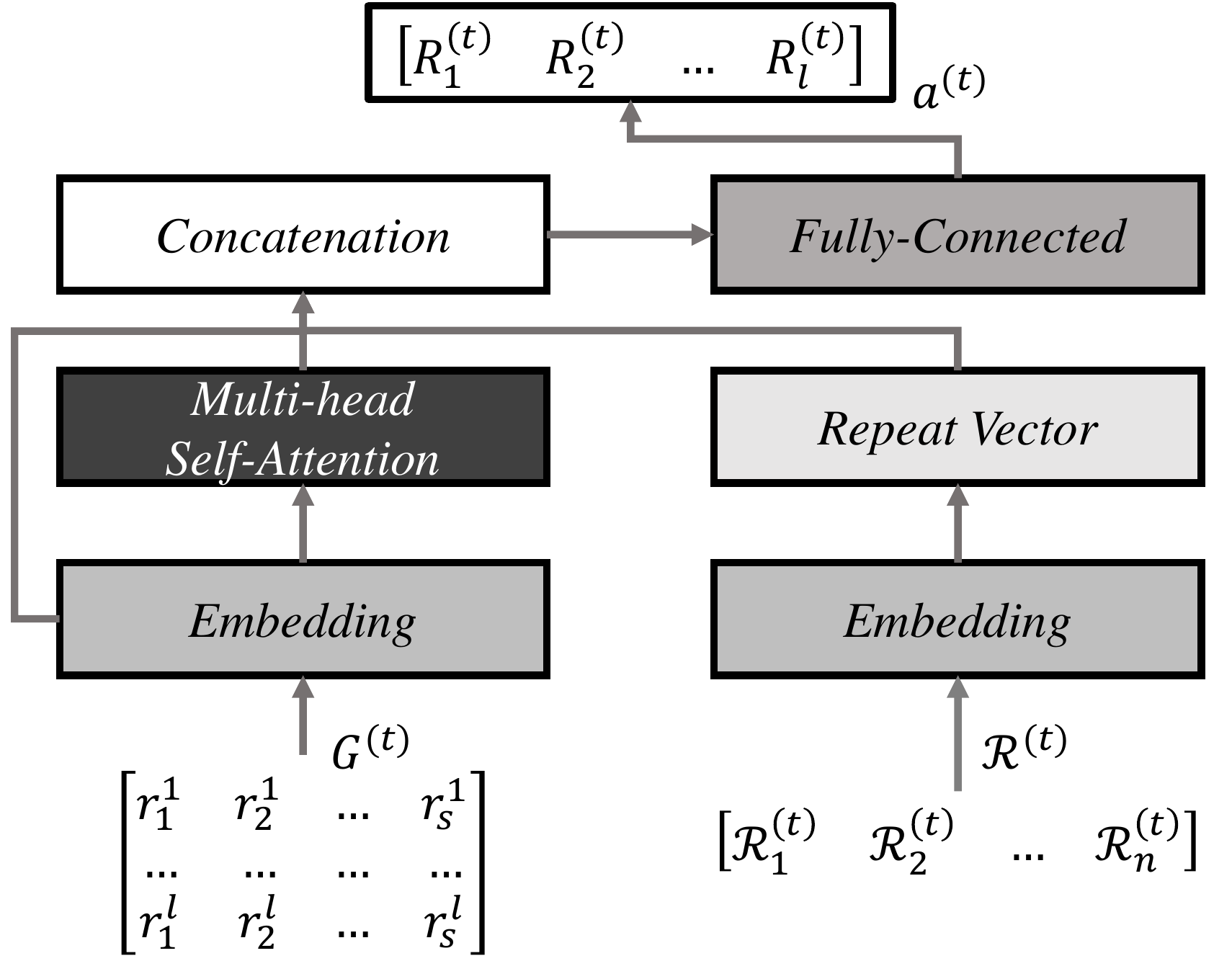}
\caption{DAA deep neural network architecture}
\vspace{-10pt}
\label{Fig:dnn}
\end{figure}

\vspace{3pt}
\noindent \textbf{Reward Network Architecture.}
In this subsection, we describe the architecture of SN. First, each input component among $\mathcal{R}^{(t)}$ and $R^{(t)}$ are fed into a different embedding network. In short, an embedding network consists of multiple fully-connected layers which transform an input vector into embedding vectors, usually of lower dimensionality. Let the mappings represented by the embedding networks for $\mathcal{R}^{(t)}$ and $R^{(t)}$ be $Emb_\mathcal{R}(\cdot)$ and $Emb_R(\cdot)$, respectively, then
\begin{equation}
\begin{split}
    Emb_\mathcal{R}(\mathcal{R}^{(t)}) = U^{(t)}_\mathcal{R} = \begin{bmatrix}
        u^{\mathcal{R}_{(t)}}_1 & u^{\mathcal{R}_{(t)}}_2 &\dots & u^{\mathcal{R}_{(t)}}_{e_\mathcal{R}}
    \end{bmatrix} \\
    Emb_R(R^{(t)}) = U^{(t)}_R = \begin{bmatrix}
        u^{(t)}_{1,1} & u^{(t)}_{1,2} & \dots & u^{(t)}_{1,e_R} \\
        \dots & \dots & \dots & \dots \\
        u^{(t)}_{l,1} & u^{(t)}_{l,2} & \dots & u^{(t)}_{l,e_R} \\
    \end{bmatrix}
\end{split}
\end{equation}
where $e_\mathcal{R}$ is the size of the output embedding for $\mathcal{R}^{(t)}$, and $e_R$ is the size of the output embedding for $R^{(t)}$.

For DRN to further learn the interrelationships among all the pending slices, $U^{(t)}_R$ is input into a \textit{Multi-Head Self-Attention} (MHSA) block \cite{vaswani2017attention}. In short, a MHSA block outputs a "context" matrix in which each row represents the "context score" of a slice given the rests in $R^{(t)}$. The MHSA block allows DAA to further analyze all pending slice requests while selecting the optimal one to accommodate. Let the mapping by the MHSA block be $S^{(t)}_R$, then
\begin{equation}
    S^{(t)}_R = \begin{bmatrix}
        \sigma^{{(t)}}_{1,1} & \dots & \sigma^{{(t)}}_{1,l} \\
        \dots & \dots & \dots \\
        \sigma^{{(t)}}_{l,1} & \dots & \sigma^{{(t)}}_{l,l}
    \end{bmatrix}
    \vspace{-5pt}
\end{equation}
where $\sigma^{{(t)}}_{i,j}$ is the context score of $G^i$ with respect to slice $G^j$. We repeat $u_{\mathcal{R}^{(t)}}$ $l$ times, then concatenate them with the rows in $U^{(t)}_R$ and $S^{(t)}_R$ to form the embedding for each slice:
\begin{equation}
\small
    U^{(t)} =
    \begin{bmatrix}
         u^{\mathcal{R}_{(t)}}_1 ... &  u^{\mathcal{R}_{(t)}}_{e_\mathcal{R}} & u^{(t)}_{1,1} ... & u^{(t)}_{1,e_R} & \sigma^{{(t)}}_{1,1} ... & \sigma^{{(t)}}_{1,l} \\
         \dots & \dots & \dots & \dots & \dots & \dots\\
         u^{\mathcal{R}_{(t)}}_1 ... & u^{\mathcal{R}_{(t)}}_{e_\mathcal{R}} & u^{(t)}_{l,1} ... & u^{(t)}_{l,e_R} & \sigma^{{(t)}}_{l,1} ... & \sigma^{{(t)}}_{l,l}
    \end{bmatrix}
\end{equation}
Finally, $U$ is input into a block of fully-connected layers that output a vector of $|\mathcal{D}|$ values $r^{(t)}_1, r^{(t)}_2, \dots, r^{(t)}_{|\mathcal{D}|}$ that represent the reward if DAA select each slice. The architecture of DAA's deep neural network is shown in Fig. \ref{Fig:dnn}.

\vspace{3pt}
\noindent \textbf{Training Algorithm.}
We train DAA deep neural network as a supervised deep neural network. More specifically, DRN is trained as a supervised model to predict the true reward of taking an action. In an episode, we first randomize the slice set $G$. Then, for each step in an episode, the input states are fed to the model to generate actions then updated for the next step as described in Section $\S$\ref{sec:method}-B. At the end of an episode, the reward values for each step are computed backward using equation (\ref{eq:reward}).

In terms of training objective, we utilize \textit{Mean Squared Error} function. Because the model only knows the reward of actions that it has taken, we utilize a \textit{mask} vector $m^{(t)}$ of size $l$ in which $m^{(t)}_i = 1$ if slice $i$ is selected in $a^{(t)}$, and $0$ otherwise. $m^{(t)}$ prevents non-selected slices from contributing to the loss value. The loss for one episode $\mathcal{E}_i$ is computed as
\begin{equation}
    \mathcal{L}_i = \sum_{t \in \text{episode}}(\hat{P}^{(t)}*m^{(t)} - P^{(t)}m^{(t)})^2
    \vspace{-5pt}
\end{equation}
where $\hat{P}$ is the predicted reward vector that is output by DRN, and $*$ represents an element-wise multiplication.

To improve performance of the training process, we further apply the Experience Relay strategy. More specifically, after being used, data (input states and output rewards) of an episode is stored in a memory dataset. After the memory has more than $n_b$ episodes, with $n_b$ being the training batch size, $(n_b - 1)$ among the generated episodes are randomly sampled and combine with the current episode to train DRN. Finally, the loss of the batch is averaged across its episodes.

\section{Experiment and Evaluation}
\label{sec:exp}

We extensively test DAA in slicing a substrate network in different conditions from easy to more difficult. More specifically, we start from substrate networks of 100 nodes of which available resources are randomized in $[10,30]$ and averages at $20$, and a slice set of $20$ slices each of which consists of $10$ VNFs with demands randomized in $[1,19]$ and averaged at $10$. As both the available resources and demands total at $2,000$, this is the case where resources are plenty. Then, we test DAA in use cases that resources are scarcer compared to the demands from VNFs in the slices that need accommodations:
\begin{enumerate}
    \item Number of slices increases from $20$ to $25$
    \item Number of VNFs per slice increases from $10$ to $15$
    \item Amount of demanding resources from the VNFs increases, from averaged at $10$ to averaged at $15$
    \item Available resources in the substrate nodes decrease, from $20$ to $15$
\end{enumerate}

After fine-tuning, the selected deep network architectures are as 1) all embedding blocks have three fully-connected layers; each layer in the substrate embedding block has a size of $n$, and each layer in the slice embedding block has a size of $s$. 2) the MHSA has five attention heads of each has a size of $s$. And 3) the fully-connected block that outputs return values has three layers, each has $3(n+s)$ neurons. The hyper-parameters $\alpha$, $\beta$, and $\lambda$, for the reward function in Eq. (\ref{eq:reward}) are set at $0.2$, $-1$, and $0.9$, respectively. The learning rate $l_r$ for training DRN as in Eq. (\ref{eq:reward}) is $0.1$. Finally, the mini-batch size in all experiments, $n_b$ is $256$. In all cases, DRN is trained with ADAM optimization algorithm with learning rate of $10^{-6}$ for $500$ epochs.

\begin{figure}[tb]
\center
    \subfigure[Increasing slices]{\includegraphics[width=0.4\columnwidth]{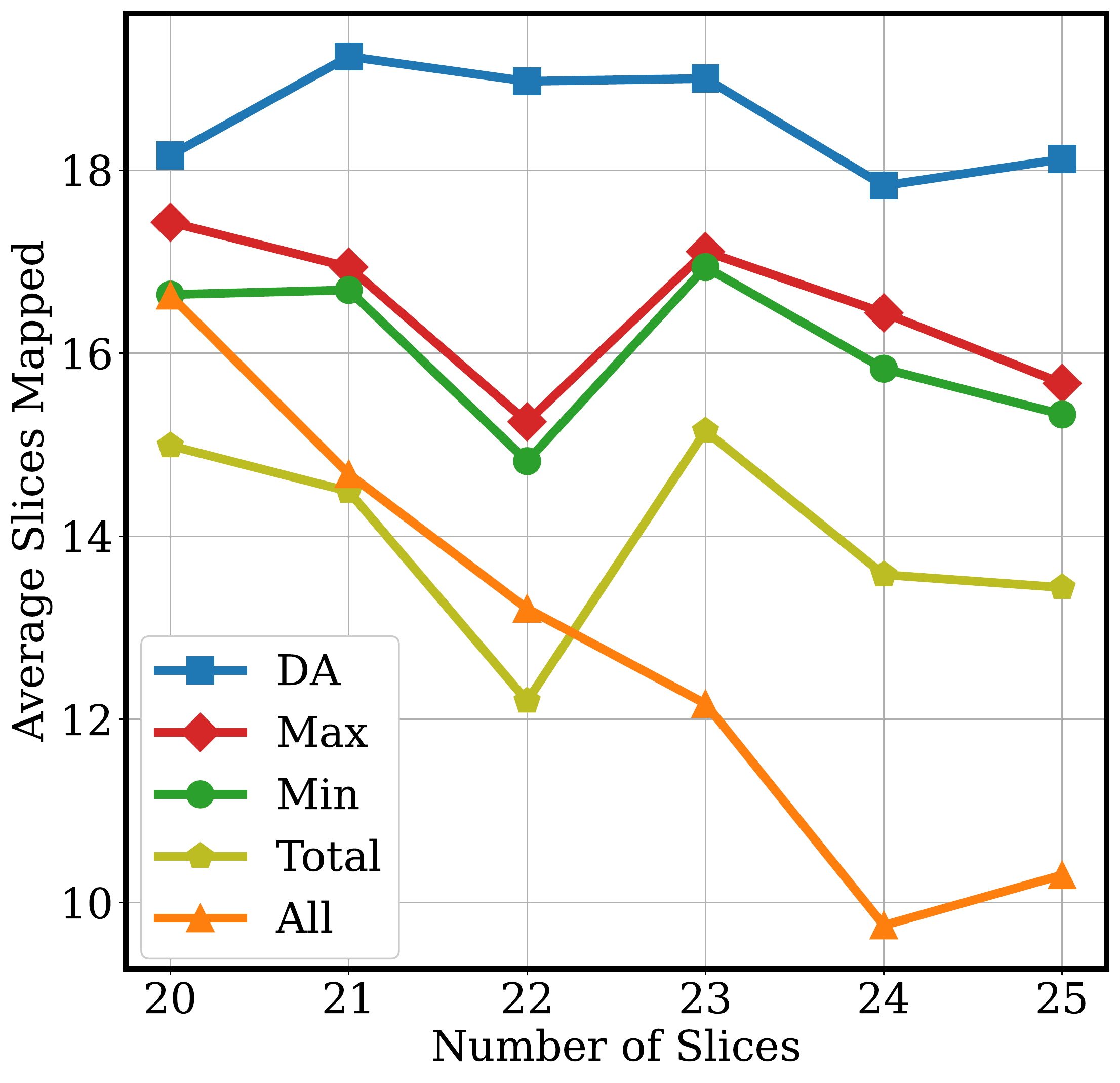}}
    \subfigure[Increasing VNFs per slice]{\includegraphics[width=0.4\columnwidth]{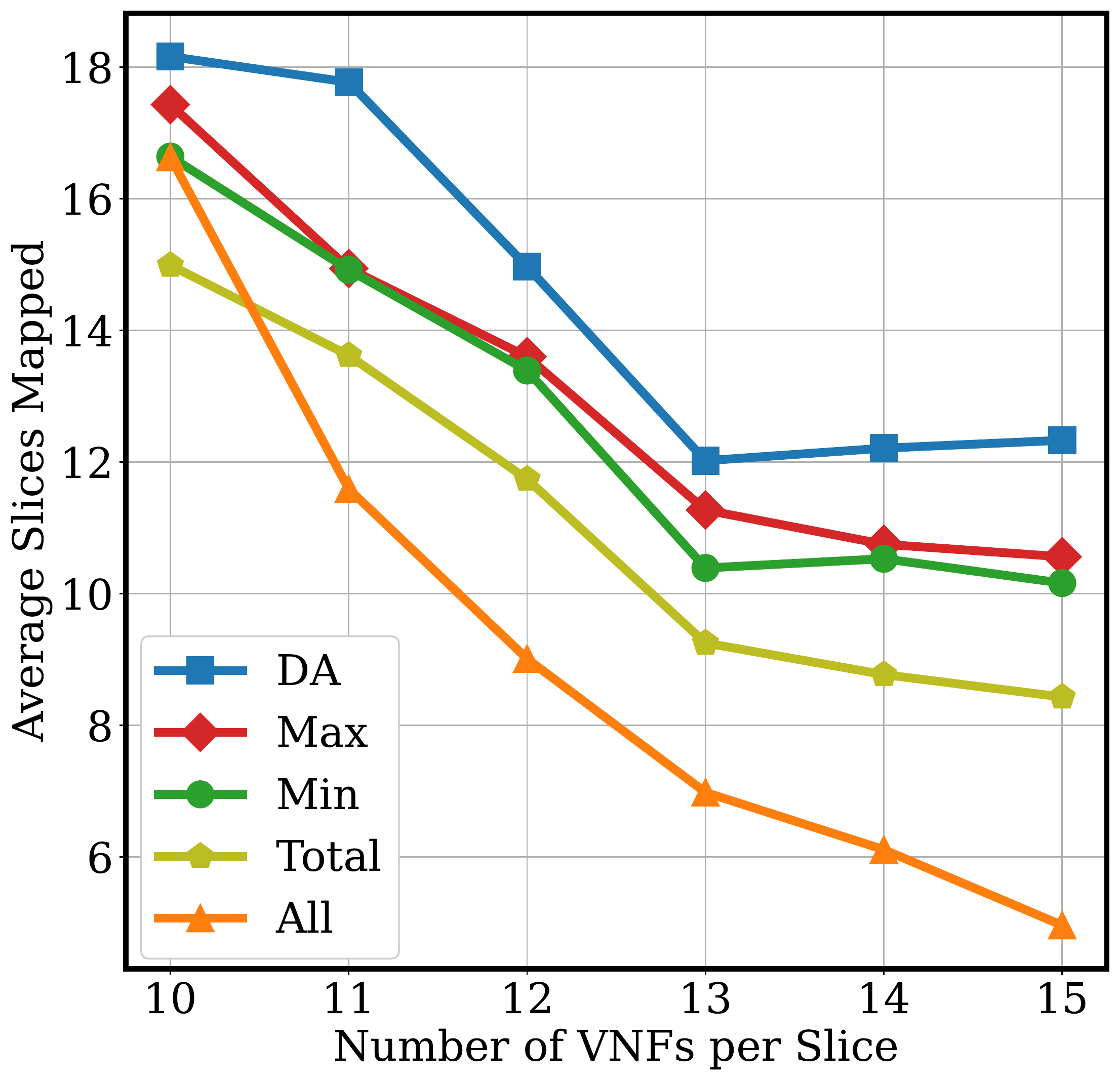}}
    \subfigure[Increasing VNF demands]{\includegraphics[width=0.4\columnwidth]{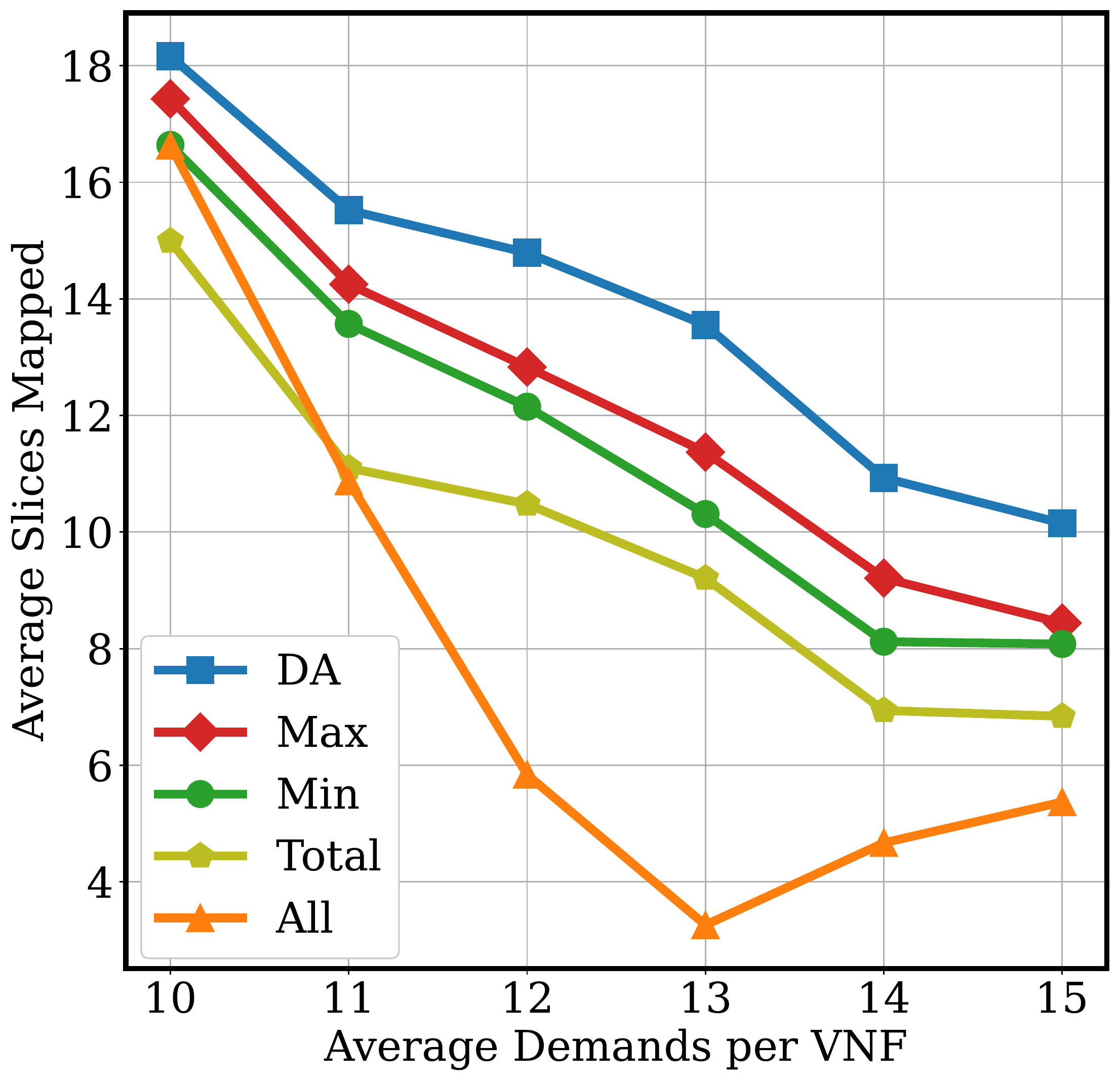}}
    \subfigure[Decreasing sub. resources]{\includegraphics[width=0.4\columnwidth]{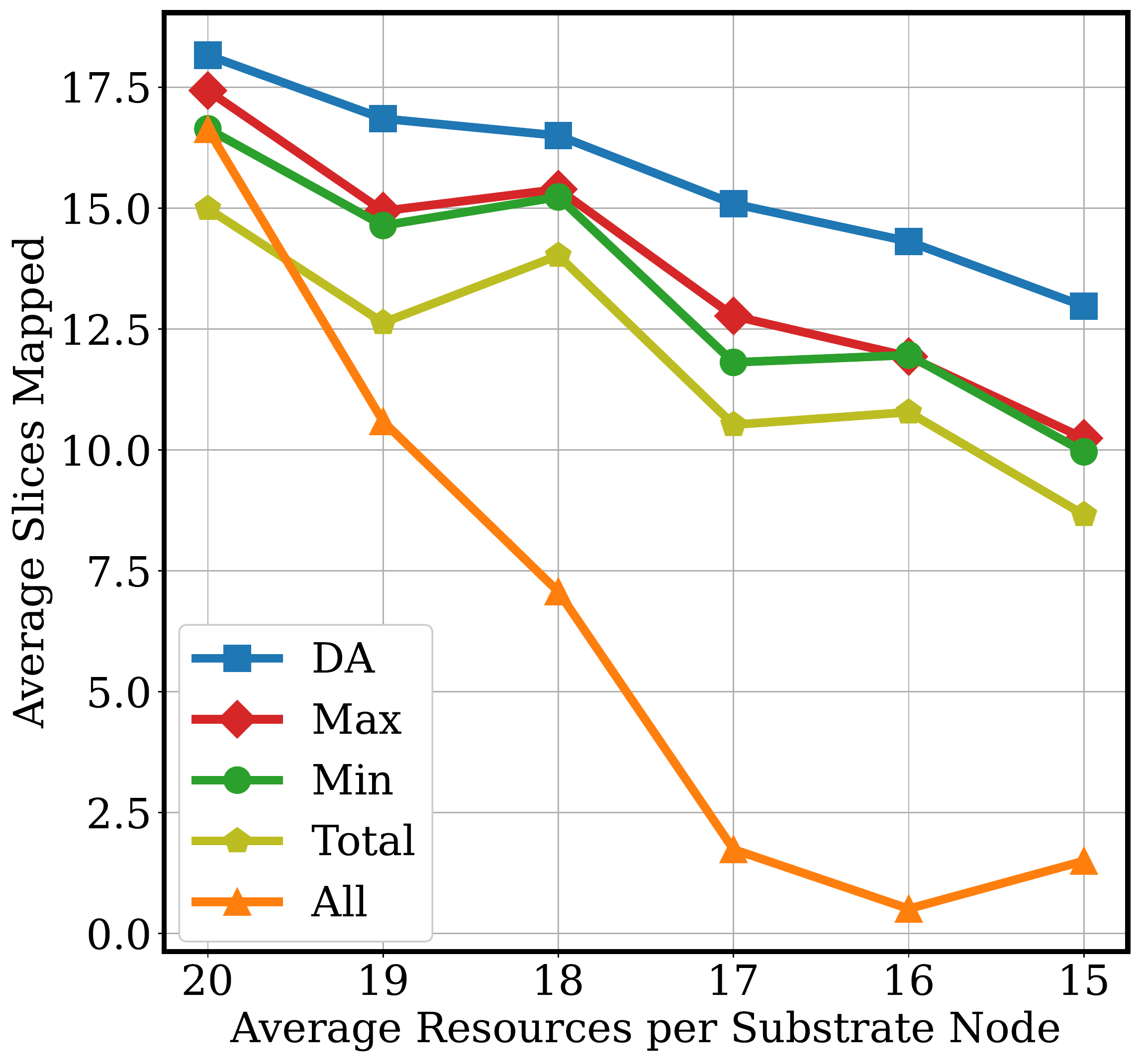}}
\vspace{-10pt}
\caption{Models' performance on  $n_G$ slices in grid networks of size $n_G \times n_G$ nodes of qubit capacity $c_i$}
\vspace{-20pt}
\label{Fig:perf}
\end{figure}

We implement four baseline strategies to compare with DAA. The first (All) performs VNF mapping purely based on their demands: all VNFs across all slices are sorted in descending order then allocated to substrate nodes using the algorithm in Section $\S$3A. The second, third, and forth baselines sort the slices in descending order of their VNFs' max demands (Max), min demands (Min), and total demands (Total). Then, in the currently selected slice, the VNFs are embedded in descending order of their individual demands. We use the average number of accommodated slices across 100 testing episodes as the evaluation metric. The are shown in Fig. \ref{Fig:perf}. As shown in Fig. \ref{Fig:perf}, in all experiment settings, DAA yields an accommodation rate that is significantly higher than that of all the naive baselines. In Fig. \ref{Fig:perf}(a) DAA maintains an accommodation rate of over $18$ in all cases, whereas the other four's performances drop gradually. In Fig. \ref{Fig:perf}(b) and (c), the decreasing patterns of all models are fairly similar (except for the All baseline); however, DAA still outperforms the others. In Fig. \ref{Fig:perf}(d), we can see DAA has the lowest decreasing rate while still yielding the highest performances.

\section{Conclusion} \label{sec:conc}

This paper addresses a fundamental challenge in RAN slicing: how to enhance the embedding performance of RAN slicing
in terms of mapping VNFs while concurrently meeting the variation of resource demands and
requirement of the RAN slices.
In terms of architecture, DAA consists of two components, an empirically designed deep neural network that was used to observe the current input states to decide the  schedule, and VNFs of the selected slice were then determined by an allocation algorithm that was aware of the substrate network's flexibility and the VNFs' allocability. We further trained the deep network of DAA as a deep reward network to predict reward the agent took after selecting and accommodating a slice. Experiment results show that, on average, DAA is able to maintain a fulfilling rate at above 80\% in a resource-limited network, and about 60\% in extreme conditions, i.e., insufficient resources at the substrate.

\section{Applications and Beyond} \label{sec:dis}
With the rapid growth of new services and Internet applications, especially massive Internet of Things, traditional cellular networks are now faced with a major challenge of supporting applications with multitude of connections such as between drones and land vehicles for smart mobility management. From a technical standpoint, the new paradigm proposed in this research offers promising new approaches for developing stable mapping plans for VNFs that are \textit{central to any drone-mounted infrastructures and systems}, with enhanced capabilities to satisfy the increasing recoverability and stability demands of services and Internet applications, while meeting their service level objectives under both normal operations and abnormal operations with failures in the network. In addition, this research provides new \textit{AI and ML based architectures} for dynamic operation in drone communications using network slicing technologies.
The proposed work is likely to have an immediate impact and will strengthen the development of applications with multitude of connections-assisted wireless communications for 5G and beyond. 

\section*{ACKNOWLEDGEMENT}
This research was supported in part by the US NSF grant CNS-2103405.

\vspace{-5pt}
\normalem
\bibliographystyle{IEEEtran}
\bibliography{bib-tn}
\end{document}